\documentclass[final,5p,times,twocolumn]{elsarticleMod}
\usepackage{amsmath,amsfonts,amssymb,array}
\usepackage{bm,bbm,bbm,bbold}
\usepackage{datetime,dsfont}
\usepackage{enumerate}
\usepackage{float}
\usepackage{graphicx,caption}
\usepackage{lineno,longtable}
\usepackage{multirow}
\usepackage{slashed,subfigure}
\usepackage{ulem}
\usepackage[svgnames]{xcolor}
\usepackage{xspace}
\usepackage[colorlinks = true,linkcolor = blue]{hyperref}
\usepackage{mydefs}
\biboptions{sort&compress}
\usepackage{tikz}
\DeclareRobustCommand{\swatch}[1]{\tikz[baseline=-0.6ex]\node[fill=#1,shape=rectangle,draw=black,thick,minimum width=5mm,rounded corners=0.5pt](){};}
\newcommand{\met}{\ensuremath{E_T^{\rm miss}}}

\hypersetup{colorlinks=true, linkcolor=blue, citecolor=ForestGreen,
filecolor=magenta, urlcolor=ForestGreen,
  pdftitle={Butterworth, et al, [arXiv:2210.13496]},
  pdfauthor={Butterworth, et al},
  pdfsubject={Subject},pdfcreator={Creator},pdfproducer={Producer},
  pdfkeywords={Type II Seesaw}{W boson mass}{Large Hadron Collider}{Electroweak Physics}
}

\graphicspath{{Figs/}}

\newcommand{\orcid}[1]{\,\href{https://orcid.org/#1}{\includegraphics[width=9pt]{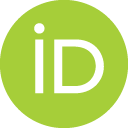}}}
\newcommand{\orcidRR}{0000-0002-3316-2175} 
\newcommand{\orcidJB}{0000-0002-5905-5394} 
\newcommand{\orcidSJ}{0000-0003-1208-6940} 
\newcommand{\orcidJH}{0000-0003-2653-5962} 
\newcommand{\orcidOM}{0000-0002-7302-7744} 
 
\newcommand{\invfb}{{\rm ~fb^{-1}}}

\def\eV{{\rm\ eV}}
\def\keV{{\rm\ keV}}
\def\MeV{{\rm\ MeV}}
\def\GeV{{\rm\ GeV}}
\def\TeV{{\rm\ TeV}}

\definecolor{magenta}{HTML}{FF00FF}
\definecolor{cornflowerblue}{HTML}{6495ED}
\definecolor{turquoise}{HTML}{40E0D0}

\definecolor{darkgreen}{rgb}{0.0, 0.2, 0.13}
\definecolor{darkmagenta}{rgb}{0.55, 0.0, 0.55}
\definecolor{amber}{rgb}{1.0, 0.6, 0.0}

\journal{Preprint numbers: IRMP-CP3-22-47, IFJPAN-IV-2022-15,MCNET-22} 
\begin{document}
\begin{frontmatter}
\title{Testing the Scalar Triplet Solution to CDF's Fat $W$ Problem at the LHC}

\author[UCL]{Jon Butterworth\ \orcid{\orcidJB}}
\ead{J.Butterworth@ucl.ac.uk}
\author[UVA]{Julian Heeck\ \orcid{\orcidJH}}
\ead{heeck@virginia.edu}
\author[SNU]{Si Hyun Jeon\ \orcid{\orcidSJ}}
\ead{si.hyun.jeon@cern.ch}
\author[CP3]{Olivier Mattelaer \orcid{\orcidOM}}
\ead{olivier.mattelaer@uclouvain.be}
\author[IFJ]{Richard Ruiz\ \orcid{\orcidRR}}
\ead{rruiz@ifj.edu.pl}

\address[UCL]{Department of Physics and Astronomy, University College London, Gower St., London, WC1E 6BT, UK}
\address[UVA]{Department of Physics, University of Virginia,
Charlottesville, Virginia 22904-4714, USA}
\address[SNU]{Department of Physics and Astronomy, Seoul National University, 1 Gwanak-Ro, Gwanak-gu, Seoul, 08826, Korea}
\address[CP3]{Centre for Cosmology, Particle Physics and Phenomenology {\rm (CP3)}, 
Universit\'e catholique de Louvain, Chemin du Cyclotron, Louvain-la-Neuve, B-1348, Belgium}
\address[IFJ]{Institute of Nuclear Physics -- Polish Academy of Sciences {\rm (IFJ PAN)}, ul. Radzikowskiego, Krak{\'o}w, 31-342, Poland}

\begin{abstract}
The Type II Seesaw model remains a popular and viable explanation 
of neutrino masses and mixing angles. 
By hypothesizing the existence of a scalar that is a triplet under the weak gauge interaction, 
the model  predicts strong correlations among neutrino oscillation parameters, signals at lepton flavor experiments, and collider observables at high energies. 
We investigate reports that the Type II Seesaw can naturally accommodate recent measurements by the CDF collaboration, which finds the mass of the $W$ boson to be significantly larger than allowed by electroweak precision data, while simultaneously evading constraints from direct searches.
Experimental scrutiny of this parameter space  in the Type II Seesaw has  long been evaded since it is not characterized by ``golden channels'' at colliders but instead by cascade decays, moderate mass splittings, and many soft final states.
In this work, we test this parameter space against publicly released measurements made at the Large Hadron Collider.
By employing a newly developed tool chain combining 
\sw{MadGraph5\_aMC@NLO}
and \sw{Contur}, we find that most of the favored space for this discrepancy is already excluded by measurements of Standard Model final states.
We give suggestions for further exploration at Run III of the LHC, which is now underway.
\end{abstract}


\end{frontmatter}


\section{Introduction}\label{sec:Intro}

In April 2022, the CDF collaboration at Fermilab reported its legacy measurement of the $W$ boson's mass $(M_W)$ using $\mathcal{L}=8.8\invfb$ of $p\overline{p}$ collision data at $\sqrt{s}=1.96\TeV$ from the Tevatron using the so-called template method~\cite{CDF:2022hxs}.
At a value of
\begin{align}
\label{eq:mw_cdf}
 M_W^{\rm CDF} = 80.4335\GeV\pm9.4\MeV\ ,   
\end{align}
this precision measurement exceeds by many standard deviations both the presently accepted 
LEP2+Tevatron average of~\cite{deBlas:2016ojx} 

\begin{align}
 M_{W}^{\rm LEP2+Tev.}=80.385\GeV\pm 15\MeV   
\end{align}
and the LEP2+Tevatron+LHC\footnote{The LHC contribution is solely from the ATLAS collaboration's Run I measurement~\cite{ATLAS:2017rzl} and does not include LHCb's early Run II measurement~\cite{LHCb:2021bjt}. 
The CMS collaboration has not yet reported a measurement of $M_W$.} ``world average''
of~\cite{deBlas:2021wap}

\begin{align}
M_{W}^{\rm World~2021}=80.379\GeV\pm12\MeV \ ,    
\end{align}
which are predicted in the Standard Model (SM) by electroweak (EW) precision data (EWPD).
Importantly, studies so far show that improvements in parton density functions~\cite{Gao:2022wxk} and perturbative matrix elements~\cite{Martin:2022qiv,Isaacson:2022rts} cannot account for this large discrepancy.
However, some SM explanations remain unexplored.
For example: the difference may be due to high-twist power corrections that are normally neglected in perturbative calculations~\cite{Ellis:1982wd,Ellis:1982cd,Collins:2011zzd}.
Alternatively, CDF's measurement of $M_W$ is ultimately a one-parameter fit of the $W$ boson's transverse mass. This distribution   also depends on the $W$'s width $(\Gamma_W)$, and so a two-parameter fit of $(M_W,\Gamma_W)$ may  reveal a shift in $\Gamma_W$.

Since CDF's finding,
numerous beyond the SM (BSM) solutions with varying complexity, novelty, and tenability have been proposed.
One outlier among these scenarios is the Type~II Seesaw model for neutrino masses~\cite{Konetschny:1977bn,Magg:1980ut, Cheng:1980qt,Schechter:1980gr,Lazarides:1980nt, Mohapatra:1980yp}, which has long predicted a shift in the $W$-to-$Z$ mass ratio.
The model is characterized by the existence of 
colorless scalars $\Delta^{\pm\pm}, \Delta^\pm, \Delta^0, \xi^0$ that  
(i) form a triplet $(\hat{\Delta})$ under the weak gauge interaction, 
(ii) carry lepton number, 
and (iii) couple to EW  gauge bosons and leptons at tree level.
More relevantly, after electroweak symmetry breaking (EWSB), 
$\hat{\Delta}$ acquires a vacuum expectation value (vev) 
$v_\Delta = \sqrt{2}\langle\hat{\Delta}\rangle$ 
that reduces the $W$-to-$Z$ mass ratio at tree level~\cite{Ross:1975fq}, but  
at one loop  the mass splittings of multiplets can 
\textit{increase} the ratio, and hence increase $M_W$~\cite{Lavoura:1993nq,Chun:2012jw}.

Dedicated studies~\cite{Kanemura:2022ahw,Heeck:2022fvl,Ghosh:2022zqs,Bahl:2022gqg,Cheng:2022hbo} point to an intriguing narrative~\cite{Heeck:2022fvl}:
a vev above $1\MeV$ and scalar masses between $m_\Delta \sim 100\GeV$ and $300\GeV$ cannot only resolve CDF's contention with EWPD but also evade constraints from direct searches by the ATLAS~\cite{ATLAS:2017xqs,ATLAS:2018ceg,ATLAS:2021jol,ATLAS:2022yzd} and CMS~\cite{CMS:2012dun,CMS:2016cpz} collaborations at the Large Hadron Collider (LHC) as well as evade constraints from searches for lepton-flavor-violating decays of charged leptons.

In this study, we draw attention to the fact that the masses, mass splittings, and vev needed for this resolution give rise to a complex collider phenomenology.
In this regime, the Type II Seesaw is not characterized by so-called ``golden channels'' at colliders.
Instead,  the mass splittings and effective couplings are so large that the decays of triplet scalars are dominated by the decays to one or more (virtual or on-shell) weak bosons and lighter, unstable scalars~\cite{Han:2007bk,FileviezPerez:2008wbg,FileviezPerez:2008jbu,Melfo:2011nx,Ashanujjaman:2021txz,Mandal:2022zmy}.
While this weakens searches for triplet scalars decaying predominantly to lepton pairs or on-shell weak boson pairs, i.e., golden channels, it also leads to a multitude of processes with many final-state charged leptons, neutrinos, and jets possessing various kinematics. 
Assuming the findings of Refs.~\cite{Kanemura:2022ahw,Heeck:2022fvl,Ghosh:2022zqs,Cheng:2022hbo} are realized in nature, 
it follows that a multitude of measurements at the LHC, especially differential cross sections for multilepton production, contain contributions of some degree  from  Type II scalars.

In light of this, we have investigated the constraining power of measurements of SM signatures from the LHC on the Type II Seesaw in the preferred parameter space identified by Ref.~\cite{Heeck:2022fvl}.
We employ a tool chain that interfaces  
\MGaMCatNLO (\MGaMCatNLOShort)~\cite{Stelzer:1994ta,Alwall:2014hca} and
\CONTUR ~\cite{Buckley:2021neu,Butterworth:2016sqg}, and
which accesses the measurements from 161 publications from ATLAS, CMS, and LHCb. We use the \CONTUR mode which accesses the SM predictions directly, as discussed in \cite{Altakach:2021lkq}, and so of these papers, we use only the subset (46) for which  SM predictions are currently available in \CONTUR. 
In practice, only a handful of these analyses
drive our results, as will be discussed in the relevant section.
As a consequence, we definitively test 
a parameter space in the Type II Seesaws that, until now, escaped experimental scrutiny.  

The remainder of this study continues as follows:
In Sec.~\ref{sec:theory_model} we summarize the Type II Seesaw model, 
current experimental constraints, and best-fit parameter spaces.
In Sec.~\ref{sec:method}, we outline our methodology and the tools used.
We present our results in Sec.~\ref{sec:results}. 
In Sec.~\ref{sec:conclusions} we conclude with an outlook for future work.

\section{Theoretical framework}\label{sec:theory_model}

Throughout this study, we work in the framework of the Type~II Seesaw model for neutrino masses~\cite{Konetschny:1977bn,Magg:1980ut, Cheng:1980qt,Schechter:1980gr,Lazarides:1980nt, Mohapatra:1980yp}. 
This scenario extends the SM's field content by a single complex, scalar multiplet. In the gauge basis, this field is denoted by $\hat{\Delta}$ following the notation of Ref.~\cite{Fuks:2019clu}.
Under the SM gauge group 
$\mathcal{G}_{\rm SM}=$SU$(3)_c$ $\otimes$SU$(2)_L$ $\otimes$U$(1)_Y$, the multiplet  carries the gauge quantum numbers 
$(\mathbf{1},\mathbf{3},+1)$.
This means that the individual states comprising $\hat{\Delta}$ are colorless but carry EW charges and couple to EW bosons via gauge couplings.
The weak hypercharge and isospin charges are normalized such that the electromagnetic charge operator is $\hat{Q}=\hat{T}_L^3+\hat{Y}$.
In terms of states with definite electric charge, $\hat{\Delta}$ and its vev $(v_\Delta)$ are, respectively,
\begin{align}
\hat\Delta =   
\begin{pmatrix}
\frac{1}{\sqrt{2}} \hat\Delta^+ & \hat\Delta^{++}\\
\hat\Delta^0 & -\frac{1}{\sqrt{2}} \hat\Delta^+
\end{pmatrix}
\  
\quad \text{and}\quad 
\langle\hat\Delta\rangle = 
\frac{1}{\sqrt{2}}\ 
\begin{pmatrix}
0 & 0 \\
v_\Delta & 0 
\end{pmatrix}
\ .
\end{align}
The tree-level Lagrangian of the Type II Seesaw is given by
\begin{align}
 \mathcal{L}_{\rm Type~II} &= 
 \mathcal{L}_{\rm SM}\ + 
 \mathcal{L}_{\rm Kin.}\ +
 \mathcal{L}_{\rm Yukawa}\ +
 \mathcal{L}_{\rm Scalar}\ .
\end{align}
Here, $\mathcal{L}_{\rm SM}$ is the SM Lagrangian.
$\mathcal{L}_{\rm Kin.}$ is the kinetic term
\begin{align}
 \mathcal{L}_{\rm Kin.} &= 
 {\rm Tr}\big[D_\mu \hat\Delta^\dag D^\mu\hat\Delta\big]\,,\
 \end{align}
 wherein the covariant derivative for $\hat{\Delta}$ is 
\begin{align}
 D_\mu \hat\Delta\ =\  \partial_\mu\hat\Delta
   - \frac{i}{2} g_W W_\mu^k \big[\sigma_k\hat\Delta-\hat\Delta\sigma_k\big]
   - i g_Y B_\mu \hat\Delta \ .
\end{align}
In the above, $W_\mu^k$, with $k=1,\dots,3$, are the weak gauge states before EWSB and couple to $\hat{\Delta}$ with a universal strength of $g_W\approx0.65$.
$\sigma_k$ are the usual $2\times2$ Pauli matrices.
$B_\mu$ is the weak hypercharge gauge state that couples with a strength of 
$g_Y = e/\cos\theta_W \approx0.36$, where $\theta_W\approx29^\circ$ is the weak mixing angle.

The term $\mathcal{L}_{\rm Yukawa}$ is the Yukawa interaction between $\hat{\Delta}$ and the left-handed SM leptons, 
\begin{align}
\label{eq:lag_yukawa}
  \mathcal{L}_{\rm Yukawa} = -{\bf Y}_\Delta \bar L^c\,\hat\Delta\, L+\text{H.c.}\ ,
\end{align}
where ${\bf Y}_\Delta$ is a complex symmetric $3\times3$ matrix of Yukawa couplings.
Eq.~\eqref{eq:lag_yukawa} conserves lepton number if we assign $L_\Delta = -2$ to $\hat{\Delta}$.
After EWSB, the term generates  Majorana masses for neutrinos
and is given by the following in the flavor basis
\begin{align}
   \mathcal{L}_{\rm Yukawa}\ \ni\  \frac{v_\Delta}{\sqrt{2} }
   ({\bf Y}_\Delta)_{ff'} \ \overline {\nu_{Lf}^c}\cdot\nu_{Lf'}
     + {\rm H.c.} 
\end{align} 
The pre-factor corresponds to the neutrino mass matrix, which can be diagonalized by a unitary rotation.
Diagonalizing allows us to express the Yukawa couplings in terms of 
the diagonal neutrino mass matrix $m_\nu^{\rm diag}$ as well as the mixing angles and phases within the Pontecorvo-Maki-Nakagawa-Sakata  matrix $U^{\rm PMNS}$.
In terms of these, the Yukawa coupling matrix is
\begin{align}
{\bf Y}_\Delta = \frac{1}{\sqrt{2}v_\Delta}\left(U^{\rm PMNS}\right)^* m_\nu^{\rm diag} \left(U^{\rm PMNS}\right)^\dagger\ ,
\end{align}
The direct connection between oscillation parameters and Yukawa couplings implies that the decays of $\hat{\Delta}$'s components to leptons at high-energy colliders are correlated with neutrino oscillation data~\cite{FileviezPerez:2008wbg,FileviezPerez:2008jbu}.
For predictions of correlations using up-to-date oscillation data, see Refs.~\cite{Fuks:2019clu,Mandal:2022zmy}.
Notice that even with full knowledge of neutrino masses and mixing parameters, the overall scale of ${\bf Y}_\Delta$ is degenerate with the magnitude of $v_\Delta$.

Finally, the scalar potential $\mathcal{L}_{\rm Scalar}$,
which includes the self-couplings of $\hat\Delta$ and
couplings of $\hat{\Delta}$ to the SM Higgs doublet $\varphi$ before EWSB is 
\begin{align}
  -\mathcal{L}_{\rm Scalar} = &\
     m_{\hat{\Delta}}^2 {\rm Tr}\left[{\hat\Delta^\dagger \hat\Delta}\right]
    + \lambda_{\Delta 1} \left({\rm Tr}\left[{\hat\Delta^\dagger \hat\Delta}\right]\right)^2
    + \lambda_{\Delta 2} {\rm Tr}\left[{(\hat\Delta^\dagger \hat\Delta)^2}\right] \nonumber\\
  &\ + \lambda_{h \Delta 1}\ \left(\varphi^\dagger \varphi \right)\ 
          {\rm Tr}\left[{\hat\Delta^\dagger\hat\Delta}\right]
     + \lambda_{h \Delta 2}\ \left(\varphi^\dagger \hat\Delta\hat\Delta^\dagger\varphi\right)
  \nonumber\\&\ 
     + \mu_{h\Delta} \Big(\varphi^\dagger \hat\Delta \cdot \varphi^\dagger +
        \text{H.c.} \big)\ .
        \label{eq:lag_scalar}
\end{align}
The parameters $\{\lambda\}$ are real dimensionless couplings.
The dimensionful parameter $\mu_{h\Delta}$ 
signifies the scale below which lepton number is not  conserved.
Under canonical quantum number assignments, SM leptons and antileptons carry lepton numbers $L_{\rm lep.}=+1$ and $L_{\rm antilep.}=-1$, respectively, 
the SM Higgs carries no lepton number,
and  $\hat{\Delta}$ carries $L_\Delta = -2$. 
 
The $\varphi-\varphi-\Delta$ term in the third line of Eq.~\eqref{eq:lag_scalar}, therefore, breaks lepton number symmetry explicitly by two units and induces the following vev for $\hat{\Delta}$ after EWSB:
\begin{align}
\sqrt{2}\langle\hat{\Delta}\rangle= 
v_\Delta\approx  \frac{\mu_{h\Delta}v_{\varphi}^2}{\sqrt{2}m_{\hat{\Delta}}^2}
 \ .
\end{align}
Here, $v_{\varphi} = \sqrt{2}\langle\varphi\rangle$ is the vev of the SM Higgs 
and we assume  
$m_{\hat{\Delta}} \gg v_{\varphi}$.
Measurements of $M_W$ in conjunction with the absence of flavor-violating decays of $\tau$ and $\mu$ leptons require $v_\Delta \gtrsim 10\eV-1\keV$,  depending on one's underlying 
assumptions~\cite{Pich:1984uoh,Chakrabortty:2012vp,Cheng:2022jyi,Heeck:2022fvl}.
Constraints on $v_\Delta$ will be revisited in Sec.~\ref{sec:theory_wmass}.

The mass eigenstates of the Type II Seesaw are those of the SM, with 
at least two
massless neutrinos being replaced by Majorana neutrinos, plus one doubly charged scalar $\Delta^{\pm\pm}$, one singly charged scalar $\Delta^\pm$, one neutral CP-even scalar $\Delta^0$, and one neutral CP-odd scalar $\xi^0$.
For $v_\Delta \ll v_{\varphi}$, the  mass eigenvalues of these scalars obey the sum rule
\begin{align}
\delta M^2_\Delta\ \equiv\ 
M_{\Delta^{0}}^2 - M_{\Delta^{\pm}}^2 = M_{\Delta^{\pm}}^2 - M_{\Delta^{\pm\pm}}^2 = \lambda_{h \Delta 2} \frac{v_{\varphi}^2}{4}\ ,
\label{eq:sum_rule}
\end{align}
with $M_{\xi^{0}}\simeq M_{\Delta^{0}}$. Notice that $\lambda_{h \Delta 2}$ can have either sign, so the triplet hierarchy is not fixed \textit{a priori}: $\Delta^{\pm\pm}$ could be the heaviest or lightest of the components.
The new states give rise to a variety of testable predictions at hadron
colliders~\cite{Akeroyd:2005gt,delAguila:2008cj,FileviezPerez:2008jbu,FileviezPerez:2009hdc,Akeroyd:2011zza,Melfo:2011nx,Nemevsek:2016enw,Cai:2017mow}.
For recent LHC updates,
see Refs.~\cite{Cai:2017mow,Fuks:2019clu,Padhan:2019jlc,Gluza:2020qrt,Ashanujjaman:2021txz,Mandal:2022zmy,CMS:2022msk}.

At the LHC, direct searches for Type II scalars set stringent constraints but depend on benchmark assumptions and signatures~\cite{ATLAS:2017xqs,ATLAS:2018ceg,ATLAS:2021jol,ATLAS:2022yzd,CMS:2012dun,CMS:2016cpz}.
For small $v_\Delta$, the Yukawa couplings are sizable and the doubly charged scalar decays solely to charged leptons. Searches for these golden channels by ATLAS at the 95\% confidence level (CL)  using Run II data exclude  $M_{\Delta^{\pm\pm}}< 1080\GeV$~\cite{ATLAS:2022yzd}, rendering the triplet too heavy to resolve the $W$-mass anomaly.
For larger vevs, the dilepton channels become subdominant compared to the more-involved bosonic decays.
In the pair and associated production channels
\begin{subequations}
\begin{align}
    pp &\to \Delta^{++}\Delta^{--}\to W^+W^+W^-W^-\ ,
    \\
    pp &\to \Delta^{\pm\pm}\Delta^{\mp} \to W^\pm W^\pm W^\mp Z \ ,
\end{align}
\end{subequations}
and for fully leptonic {and semileptonic} final states, 
doubly charged scalar masses in the range $M_{\Delta^{\pm\pm}} = 200\GeV-350\GeV$ are excluded by ATLAS at the 95\% CL using the full Run II data set~\cite{ATLAS:2018ceg,ATLAS:2021jol}.
This exclusion assumes $v_\Delta=0.1\GeV$ but 
holds for larger $v_\Delta$.
The limits of Ref.~\cite{ATLAS:2021jol} 
are driven by searches for $\Delta^{++}\Delta^{--}$ pair production; 
limits from searches for $\Delta^{\pm\pm}\Delta^\mp$ only exclude 
$M_{\Delta^{\pm\pm}} < 230\GeV$ at 95\% CL for $\vert  M_{\Delta^\pm}-M_{\Delta^{\pm\pm}}\vert < 5\GeV$.
A recasting of 
Run I results finds $M_{\Delta^{\pm\pm}} < 84\GeV$ excluded~\cite{Kanemura:2013vxa,Kanemura:2014goa,Kanemura:2014ipa}, leaving an allowed region of  $84\GeV < M_{\Delta^{\pm\pm}} < 200\GeV$ that is suited to explain the CDF result over a large range of triplet vevs~\cite{Heeck:2022fvl,Cheng:2022hbo}.

For intermediate $v_\Delta$, it is possible for light triplet scalars to be so long-lived that they circumvent the prompt-decay limits above.
Investigations~\cite{BhupalDev:2018tox,Antusch:2018svb} find that this occurs 
roughly in the triangle 
in $(M_{\Delta^{\pm\pm}},v_\Delta)$-space
enclosed by the points:
\begin{subequations}
\label{eq:long_lived_boundary}
\begin{align}
 (M_{\Delta^{\pm\pm}},v_\Delta)&=(90\GeV,10^{-1}\GeV)\ ,
 \\
 (M_{\Delta^{\pm\pm}},v_\Delta)&=(90\GeV,10^{-4}\GeV)\ ,
 \\
 (M_{\Delta^{\pm\pm}},v_\Delta)&=(200\GeV,10^{-4}\GeV)\ .
\end{align}
\end{subequations}
For larger $(M_{\Delta^{\pm\pm}},v_\Delta)$, $\Delta^{\pm\pm}$ has a characteristic lifetime below the threshold $c\tau_0=1$ mm, and is therefore short lived.

We briefly note that the search for Higgs pair  production at ATLAS with its full Run II data set~\cite{ATLAS:2021tyg} has been reinterpreted for the Type II Seesaw. Specifically, Ref.~\cite{Bahl:2022gqg} reports that the absence of an enhanced, loop-induced $H\to\gamma\gamma$  rate in this search leads to  $M_{\Delta^{\pm\pm}}\lesssim250\GeV$ being excluded.
The values of $(M_{\Delta^{\pm\pm}},M_{\Delta^{\pm}})$ that can explain CDF's data, however, lie  along the boundary of this reinterpreted limit. This limit is also subject to several uncertainties, including corrections to Higgs pair production, which is assumed to be SM-like.

\subsection{\texorpdfstring{$M_W$}{MW} in the Type II Seesaw at tree level and one loop}
\label{sec:theory_wmass}

At tree level, a nonzero $v_\Delta$ formally leads to a  $Z$ mass that is larger than predicted in the SM. When $M_Z$ is used as an EW reference point, this manifests as a tree-level $W$ mass that is \textit{smaller} than expected in the SM. 
From both perspectives, one anticipates a smaller $M_W$-over-$M_Z$ ratio, which can be quantified by the $\rho$~\cite{Ross:1975fq} and oblique $(S,T,U)$~\cite{Peskin:1990zt,Peskin:1991sw}  parameters:
\begin{align}
    \rho_{\rm tree} = \frac{M_W^2}{M_Z^2\ \cos^2\theta_W}
    =
    1 + \alpha_{\rm EM}\ T_{\rm tree}
    = 1 - \frac{2v_\Delta^2}{v_{\varphi}^2  + 2v_\Delta^2}\ .
\end{align}
In the SM, one has $\rho_{\rm tree} =1 $ and $T_{\rm tree} =0$, with deviations in the Type II Seesaw driven at tree level by $v_\Delta$.
Qualitatively, EWPD lead to the condition that $v_\Delta \ll v_{\varphi}$,
even in light of CDF's measurement~\cite{ParticleDataGroup:2020ssz,deBlas:2022hdk,Kanemura:2022ahw}.

At one loop, 
contributions to the $W$ boson's self-energy from triplet scalars lead to a shift in its mass that scales with $\delta M^2_\Delta$. In terms of the oblique parameters $S$ and $T$,
and with $U=0$ since it is small,
such contributions can be expressed as~\cite{Peskin:1990zt,Peskin:1991sw,Maksymyk:1993zm}
\begin{align}
    M_W \approx M_W^{\rm SM} \left[1 - \frac{\alpha_{\rm EM}}{4(1-2s^2_W)}(S-2(1-s^2_W)T)\right]\ ,  
    \label{eq:MW_S_T}
\end{align}
using the shorthand $s^2_W\equiv\sin^2\theta_W$.
For heavy triplet masses, small mass splittings $(\delta M_\Delta^2/ M_{\Delta^0}^2)$, and neglecting $\mathcal{O}(v_\Delta^2/v_{\varphi}^2)$ terms, the one-loop expressions in the Type II Seesaw are~\cite{Mandal:2022zmy}
\begin{subequations}
\begin{align}
S_{\rm 1-loop} &\approx -\frac{(2-4s_W^2 + 5s_W^4)M_Z^2}{30\pi M_{\Delta^0}^2} 
+ \frac{2\delta M_\Delta^2}{3\pi M_{\Delta^0}^2}\ ,
\\
T_{\rm 1-loop} &\approx \frac{\left(\delta M_\Delta^2\right)^2}{12\pi^2 \alpha_{\rm EM} M_{\Delta^0}^2 v_{\varphi}^2}\ .
\end{align}
\end{subequations}
Here, we highlight the distinction between tree-level shifts to $M_W$, 
which are driven by the (small) triplet vev $v_\Delta$, 
and one-loop shifts,
which depend on $\delta M_\Delta^2$.
Due to this difference, it is possible that positive  one-loop corrections to $M_W$ (or $T$) can exceed the negative tree-level correction to $M_W$~\cite{Kanemura:2022ahw,Heeck:2022fvl}.

\begin{figure*}[t]
\begin{center}
\includegraphics[width=.9\textwidth]{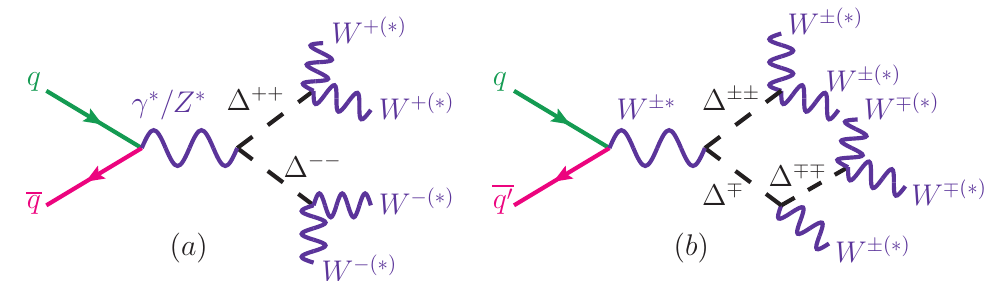} 
\end{center}
\caption{Graphs illustrating the production of Type II scalars 
in 
(a) neutral-current and (b) charged-current
quark-antiquark annihilation at the Born level
and their decays to $W$ bosons.
Drawn with \textsw{JaxoDraw}~\cite{Binosi:2008ig}.
}\label{fig:diagram}
\end{figure*}

Adjusting the $W$ mass in Eq.~\eqref{eq:MW_S_T} to match CDF's value from Eq.~\eqref{eq:mw_cdf} fixes one linear combination of $S$ and $T$.
Complementary constraints arise from EWPD, notably the weak mixing angle $\theta_W$~\cite{Bahl:2022gqg}.
A dedicated fit to EWPD including CDF's data finds the best-fit value $(S,T) = (0.17,0.27)$~\cite{Lu:2022bgw,Asadi:2022xiy}.
That $T$ is positive requires $|T_{\rm 1-loop}|>|T_{\rm tree}|$, 
which is naturally achieved for $v_\Delta \ll v$.
The positive $S$ furthermore favors $\delta M_\Delta^2>0$ within $1\sigma$, and hence favors the mass hierarchy
\begin{align}
\label{eq:theory_mass_hierarchy}
M_{\Delta^{\pm\pm}}^2\ <\ M_{\Delta^{\pm}}^2\ <\ M_{\Delta^{0}}^2\ \sim\ M_{\xi^{0}}^2\ .
\end{align}
The best fit of Ref.~\cite{Asadi:2022xiy} translates roughly to the values~\cite{Heeck:2022fvl} 
\begin{align}
\label{eq:param_best_fit}
    \left(M_{\Delta^{\pm\pm}},\ M_{\Delta^{\pm}} - M_{\Delta^{\pm\pm}}\right)\ 
    \approx\ 
    \left(95.5\GeV,\ 72.5\GeV\right)\ .
\end{align}

Using the full expressions at one loop for $S,T$~\cite{Peskin:1990zt,Peskin:1991sw} but neglecting tree-level contributions, which are $\mathcal{O}(v_\Delta^2/v_{\varphi}^2)\ll1$, Ref.~\cite{Heeck:2022fvl} finds that the uncertainty in Eq.~\eqref{eq:mw_cdf}
corresponds at $1\sigma$
to the following (correlated) parameter space that additionally remains unconstrained by direct searches for Type II Seesaw:
\begin{subequations}
\label{eq:theory_limits_global}
\begin{align}
    1\MeV \lesssim v_\Delta \lesssim 1\GeV\ ,
    \\
    84\GeV < M_{\Delta^{\pm\pm}} < 200\GeV\ ,
    \\
    56\GeV < M_{\Delta^{\pm}} - M_{\Delta^{\pm\pm}} < 81\GeV\ ,
    \\
    46\GeV < M_{\Delta^{0}} - M_{\Delta^{\pm}} < 53\GeV\ .
\end{align}
\end{subequations}

This is the preferred region of parameter space in the Type~II Seesaw  
(in the natural limit $v_\Delta \ll v_\varphi$)
that can reconcile CDF's measurement of $M_W$ with EWPD.
For larger $v_\Delta$, i.e., $v_\Delta\gtrsim 1\GeV$, one needs larger mass splittings $\delta M_\Delta^2$ in order to compensate for the larger, unwanted tree-level contribution to $T$~\cite{Kanemura:2022ahw}. In this fine-tuned region one can also have larger triplet masses than indicated by Eq.~\eqref{eq:theory_limits_global} in order to keep $S$ small. Since many of our assumptions rely on small $v_\Delta$,
which is also the more natural 
parameter space to explain CDF's finding, we restrict ourselves to 
the triplet parameter space of Eq.~\eqref{eq:theory_limits_global}.
In Sec.~\ref{sec:results}, we report a test of this parameter space by reinterpreting measurements of SM signatures at the LHC.

\section{Reinterpreting LHC measurements of SM signatures}
\label{sec:method}

The LHC detector experiments are producing a growing library of measurements of differential cross sections, exploring a wide range of particle final states in $pp$ collisions at $\sqrt{s}=7$, 8, and 13 TeV collision energies. While the primary motivation for these measurements is to probe the SM itself and test SM predictions in new kinematic regimes, many measurements are made in a sufficiently model-independent fashion that non-SM processes may be ``signal-injected'' onto  measured phase spaces. Usually denoted as a ``fiducial region,'' this restriction of phase space implements the experimental analysis cuts on the final-state particles. Thus, the acceptance (and potential discovery/exclusion) for any BSM events populating this volume of phase space can be evaluated.

To carry out this discovery/exclusion procedure, we use a new release (v2.4.0) of the tool \CONTUR.
{\CONTUR} is 
package that exploits the analysis routines published in \RIVET (v3.1.6)~\cite{Bierlich:2019rhm} and \textsw{Yoda} (v1.9.6), and which in turn obtains the measurement data from \HEPDATA~\cite{Maguire:2017ypu}.
In {\CONTUR}, a $\chi^2$ test statistic is used to compare the likelihoods that a given   measurement was obtained under competing assumptions, namely that either the SM alone or SM+BSM is the underlying distribution. The likelihood ratio is then used to derive a CL for testing a BSM hypothesis. Correlations between the uncertainties are taken into account where available. More details on {\CONTUR} may be found in Ref.~\cite{Buckley:2021neu}.

To simulate signal processes from the Type II Seesaw at the LHC,
we use the public  \sw{TypeIISeesaw} libraries of Ref.~\cite{Fuks:2019clu}.
These libraries are an implementation of the model described in Sec.~\ref{sec:theory_model} into 
the \FEYNRULES package~\cite{Christensen:2008py,Alloul:2013bka}
and are available as a set of 
Universal \FEYNRULES Object (UFO) libraries~\cite{Degrande:2011ua}.
Parton-level matrix elements are computed 
at lowest order (LO) 
by importing the above UFO
into 
\MGaMCatNLOShortV{(v2.9.10)}.

We focus on the production and decay channels
\begin{subequations}
\label{eq:process}
    \begin{align}
        p p &\to \Delta^{++}\Delta^{--}\to W^{+ (*)}W^{+ (*)}W^{- (*)}W^{- (*)}\ \label{process:pair} ,
        \\ 
        p p &\to \Delta^{\pm\pm}\Delta^{\mp}\ \to W^{\pm (*)}W^{\pm (*)}\Delta^{\pm\pm}W^{\pm (*)}
        \to 5W^{(*)}\ \label{process:associated} , 
    \end{align}
\end{subequations}
which are shown graphically in Fig.~\ref{fig:diagram},
and which are driven by the $\Delta^{\pm\pm}\to W^{\pm(*)}W^{\pm(*)}$ and
$\Delta^{\pm}\to \Delta^{\pm\pm} W^{\mp(*)}$ decay modes.\footnote{Adding the channel $ p p \to \Delta^\pm \Delta^0 $ is computationally expensive and only strengthens the significance of our excluded region. Therefore, we neglect it in our study. We also do not consider interference with SM processes, which is anyway not present for most LHC measurements, since we are considering resonant production of $\Delta^{\pm(\pm)}$ states and signal processes with 4-to-5 $W^{(*)}$ bosons.}
The $W^{(*)}$ are allowed to split into both lepton pairs and quark pairs.

Matrix elements  are convolved with the NNPDF 3.0 LO parton distribution function (\textsw{lhaid=263400})~\cite{NNPDF:2014otw}.
Decays of triplet scalars are treated using the narrow width approximation as implemented into \MGaMCatNLOShort's {\MADSPIN} module~\cite{Artoisenet:2012st,Alwall:2014bza}.
Parton-level events are passed to \PYTHIAV{(v8.306)} \cite{Bierlich:2022pfr} 
to simulate showering and hadronization,
using the Monash 2013 tune~\cite{Skands:2014pea}.
multiparton interactions are disabled in order to speed up the simulation process.
Finally, a new interface between \MGaMCatNLOShort and \CONTUR 
passes the \HEPMC outputs of \PYTHIA 
to the \RIVET routines for 
LHC Run I and Run II measurements\footnote{The full list can be found in the \CONTUR~2.4.0 release.}

The above sequence is carried out over the parameter space 
\begin{subequations}
\begin{align}
    M_{\Delta^{\pm\pm}} &\in [60\GeV,400\GeV]\ ,
    \\
    \Delta_{M} \equiv M_{\Delta^{\pm}} - M_{\Delta^{\pm\pm}}
    &\in [35\GeV,155\GeV]\ ,
\end{align}
\end{subequations}
which covers the parameter space 
summarized in Eq.~\eqref{eq:theory_limits_global}.
Scalar masses are fixed to obey the {sum rule in Eq.~\eqref{eq:sum_rule}} and the mass hierarchy in Eq.~\eqref{eq:theory_mass_hierarchy}.
While the neutral scalars $\Delta^0,\ \xi^0$ are present in the model,
we do not include their contributions since they are heavier and have smaller production cross sections.
Additionally, we fix $v_\Delta=1\GeV$; larger values increase the  $\Delta^{\pm\pm}\to W^{(*)}W^{(*)}$ branching ratio, and therefore increase the channel's signal strength. 
We have checked  numerically that this behavior is realized in our simulations.

\begin{figure}[t]
\includegraphics[width=\columnwidth]{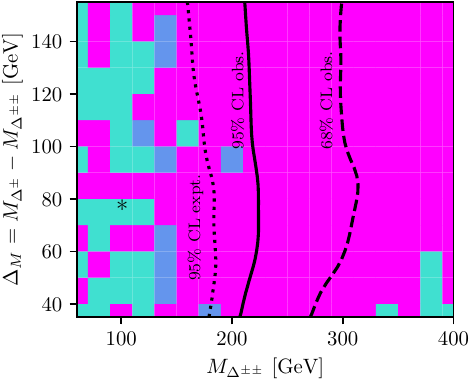} 
\begin{tabular}{ll}
        \swatch{magenta}~ATLAS 4$\ell$& JHEP,07:005, 2021\\
        \swatch{cornflowerblue}~ATLAS $\ell_1\ell_2$+\met{}& EPJC,77(3):141,2017.\\
        \swatch{turquoise}~ATLAS $\ell_1\ell_2$+\met{}+jet& JHEP,06:003, 2021.
\end{tabular}
\caption{The $(M_{\Delta^{\pm\pm}},\Delta_{M})$ parameter space overlaid with the 95\% (solid) and 68\% (long-dash) exclusion limits 
as obtained from \MGaMCatNLOShort +\ \CONTUR.
(Values to the left of the lines are excluded.)
Also shown is the 95\% expected exclusion (dotted).
The color-shading scheme indicates which SM measurement provides the dominant exclusion.
The black asterisk indicates the best fit value from \cite{Heeck:2022fvl}.
}
\label{fig:contur_plot}
\end{figure}

\section{Results}\label{sec:results}

In Fig.~\ref{fig:contur_plot}, we overlay the 
$(M_{\Delta^{\pm\pm}},\Delta_{M})$
parameter space we investigate
with exclusion limits set by \CONTUR.
As discussed in Sec.~\ref{sec:theory_model}, doubly charged scalar masses in the range $M_{\Delta^{\pm\pm}}=84\GeV-200\GeV$ are not excluded by direct searches by either ATLAS or CMS.
The best-fit point for the recent $M_{W}$ measurement from CDF, 
and summarized in Eq.~\eqref{eq:param_best_fit}, lies in 
this
window. 
We find that the limit from \CONTUR excludes all of this window.
More specifically, we report that at 95\% CL
\begin{align}
    M_{\Delta^{\pm\pm}}< 200\GeV\ \text{for}\ \Delta_M\in[35,155]\GeV
\end{align}
are excluded by measurements of SM signatures at the LHC.
We have checked that these limits extend to larger mass splittings as well, i.e., where $\Delta^\pm$ becomes irrelevant.
These limits are obtained for a vev of $v_\Delta=1\GeV$ and are applicable for larger $v_\Delta$. For smaller $v_\Delta$, 
experimental acceptance, and hence sensitivity, can degrade due to too-long scalar lifetimes.
The analyses here assume triplet scalars decay promptly after production.
For further discussion, see Sec.~\ref{sec:theory_model}.

For most of the parameter plane, the exclusion is driven by
the inclusive, four-lepton cross section measurement from ATLAS~\cite{ATLAS:2021kog}, which uses $\mathcal{L}=139\invfb$ of data at $13\TeV$. The search for triboson $WWW$ production from ATLAS at $8\TeV$ with $\mathcal{L}=20\invfb$ of data~\cite{ATLAS:2016jeu} and ATLAS' $WW$+jet measurement at $13\TeV$ with $\mathcal{L}=139\invfb$ of data~\cite{ATLAS:2021jgw} drive the exclusion for parts of the low-$M_{\Delta^{\pm\pm}}$ region.

\begin{figure}[t]
\includegraphics[width=\columnwidth]{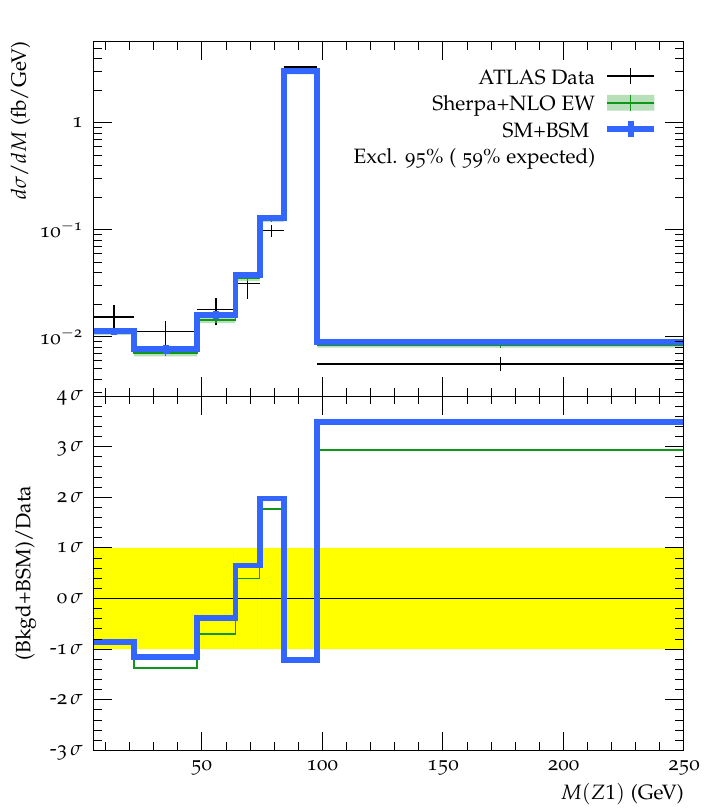} 
\caption{Upper panel:
Representative differential cross section as a function of the highest-mass dilepton pair 
in $4\ell$ measurements used in this study
showing ATLAS data (crosses)~\cite{ATLAS:2021kog}, 
predicted SM yields (green)~\cite{Sherpa:2019gpd},
and predicted SM+BSM yields 
for $(M_{\Delta^{\pm\pm}},M_{\Delta^{\pm}})=(180\GeV,255\GeV)$
(blue).
Lower panel: bin-by-bin significance of expected theory yields relative to data 
with combined data and theory uncertainties (band).
\label{fig:rivet_plot}
}
\end{figure}

To explore further the phenomenology driving the expected (dotted) and observed 95\% (solid) 
exclusion limits in  Fig.~\ref{fig:contur_plot},
we draw attention to the sensitivity to 
$M_{\Delta^{\pm\pm}}$ as a function of $\Delta_M$.
In the expected limits, there is moderate sensitivity to the mass splitting while the observed limits appear largely insensitive to $\Delta_M$. The observed limits are also somewhat more stringent than the expected limits.
The results follow from several competing and complementing factors.

Focusing first on the expected limits at small $M_{\Delta^{\pm\pm}}$ and small $\Delta_M$, 
one expects comparable signal contributions from 
the $\Delta^{++}\Delta^{--}$ pair production process in  Eq.~\eqref{process:pair}
and 
the $\Delta^{\pm\pm}\Delta^{\mp}$ associated production channel in Eq.~\eqref{process:associated}.
This follows from $\Delta^{\pm\pm}\Delta^{\mp}$ production having generically~\cite{Fuks:2019clu} a production cross section that is about $\mathcal{O}(2\times)$ larger than the
$\Delta^{++}\Delta^{--}$ production  rate combined with the $\Delta^{\pm}\to \Delta^{\pm\pm}f\overline{f'}$ decay rate reaching BR$(\Delta^{\pm}\to \Delta^{\pm\pm}f\overline{f'})\sim\mathcal{O}(50\%)$.
This means that the two channels in Eqs.~\eqref{process:pair}-\eqref{process:associated} effectively have the same cross section for small $M_{\Delta^{\pm\pm}}$ and small $\Delta_M$.

As the mass splitting increases, two competing effects in the associate production channel occur:
(i) the $\Delta^{\pm\pm}\Delta^{\mp}$ production cross section decreases due to phase space suppression,
(ii) the $\Delta^{\pm}\to \Delta^{\pm\pm}f\overline{f'}$ decay rate increases to BR$(\Delta^{\pm}\to \Delta^{\pm\pm}f\overline{f'})\gtrsim\mathcal{O}(90\%)$ for $M_{\Delta^{\pm\pm}}\gtrsim 60\GeV$ due to coupling enhancements.
The former occurs faster than the later, leading to the associated production channel effectively turning off for large $\Delta_M$,
and which translates to a decrease in expected sensitivity.

Focusing now on the observed limits, we find that the observed limits 
constrain $M_{\Delta^{\pm\pm}}$ by about $30\GeV-50\GeV$ more than the expected limits.
This comes from the fact that 
(i) the measurements we considered are statistics limited
and 
(ii) the SM prediction lies above the measured data in the most sensitive region of the measurement -- the mass distribution of each dilepton pair in the event.
An example of such a distribution is presented in Fig.~\ref{fig:rivet_plot},
which shows the mass distribution of the highest-mass dilepton pair 
in $4\ell$ events when the four-lepton system has a mass greater than twice the $Z$ mass~\cite{ATLAS:2021kog}.
Signal events for the representative mass points $(M_{\Delta^{\pm\pm}},M_{\Delta^{\pm}})=(180\GeV,255\GeV)$ have been injected.

\section{Outlook and Conclusion}\label{sec:conclusions}

Among the proposed new physics reasons for the difference between CDF's precision measurement of the $W$'s mass and the SM expectation, the Type II Seesaw model stands out for having long predicted a shift in the $W$ mass and furthermore for being a realistic scenario  to explain neutrino masses.
Recent studies~\cite{Kanemura:2022ahw,Heeck:2022fvl,Ghosh:2022zqs} indicate that in order to accommodate CDF's measurement, triplet scalars must carry masses of few-to-several hundred GeV.
This is well within the LHC's kinematic reach.
However, due to the their sizable decay rates to EW bosons and lighter triplet scalars, direct searches 
for the 
scalars  $\Delta^{\pm\pm}$ and $\Delta^{\pm}$ from the Type II Seesaw have not fully probed this window.

In this study, we exploit the fact that triplet scalars with such properties 
contribute (at some sub-leading level) to  measurements of SM signatures at the LHC. These include triboson processes, e.g., $pp\to WWW$, and diboson processes, e.g., $pp\to W/Z+nj$,
as well as agnostic searches for new physics.
Using releases of \MGaMCatNLOShort and \CONTUR that interface natively,
in conjunction with 
an updated \FEYNRULES description of the Type~II Seesaw~\cite{Fuks:2019clu},
we have performed an analysis of the Type II Seesaw using publicly  available  measurements from 
Runs I and II of the LHC. 

For promptly decaying $\Delta^{\pm\pm}$, 
we find that the best-fit point and the $1\sigma$ region in the Type II Seesaw's parameter space consistent with CDF's measurement~\cite{Heeck:2022fvl}, 
given in Eqs.~\eqref{eq:param_best_fit} and \eqref{eq:theory_limits_global} respectively,   are already excluded by LHC data at 95\% CL, 
as shown in Fig.~\ref{fig:contur_plot}.
This exclusion is driven by precision measurements of triboson and diboson processes at Run II.
This region was previously unconstrained by direct ATLAS and CMS searches.
The remaining parameter space preferred by CDF corresponds to the 
limit that triplet scalars are long-lived and prompt-decay searches do not apply, and summarized in Eq.~\eqref{eq:long_lived_boundary}. We encourage dedicated experimental searches to probe this gap.

Finally, the methodology employed here is applicable to other untested parameter spaces in the Type II Seesaw  (as well as in other models)
that feature a collider phenomenology similarly characterized by cascade decays, moderate mass splittings, and many soft final states,
i.e., not by golden collider signatures.
Since the measurements used here are generally
statistically limited, more precise measurements, with greater
kinematic coverage, should be a priority for the coming years
of high-luminosity running, and will significantly extend the model
space which can be probed by these methods.


\section*{Acknowledgements}

Benjamin Fuks, 
Jonathan Kriewald,
Miha Nemev{\v s}ek, 
and Carlo Tamtam
are thanked for helpful discussions.

The research activities of JH are supported in part by the National Science Foundation under Grant PHY-2210428. 
The work of SJ is supported by the National Research Foundation of Korea (NRF) grant funded by the Korean government (NRF-2021R1A2C2014311).
OM
received funding from FRS-FNRS agency via the IISN
maxlhc convention (4.4503.16).
RR acknowledges the support of Narodowe Centrum Nauki under Grant No. 2019/ 34/ E/ ST2/ 00186. The author also acknowledges the support of the Polska Akademia Nauk (grant agreement PAN.BFD.S.BDN. 613. 022. 2021 - PASIFIC 1, POPSICLE). This work has received funding from the European Union's Horizon 2020 research and innovation program under the Sk{\l}odowska-Curie grant agreement No.  847639 and 
as part of the Sk{\l}odowska-Curie Innovative Training Network MCnetITN3 (grant agreement no. 722104), and from the Polish Ministry of Education and Science.

\bibliography{madContur_TypeIIWMass.bib}

\begin{thebibliography}{10}
\expandafter\ifx\csname url\endcsname\relax
  \def\url#1{\texttt{#1}}\fi
\expandafter\ifx\csname urlprefix\endcsname\relax\def\urlprefix{URL }\fi
\expandafter\ifx\csname href\endcsname\relax
  \def\href#1#2{#2} \def\path#1{#1}\fi

\bibitem{CDF:2022hxs}
T.~Aaltonen, et~al., {High-precision measurement of the W boson mass with the
  CDF II detector}, Science 376~(6589) (2022) 170--176.
\newblock \href {https://doi.org/10.1126/science.abk1781}
  {\path{doi:10.1126/science.abk1781}}.

\bibitem{deBlas:2016ojx}
J.~de~Blas, M.~Ciuchini, E.~Franco, S.~Mishima, M.~Pierini, L.~Reina,
  L.~Silvestrini, {Electroweak precision observables and Higgs-boson signal
  strengths in the Standard Model and beyond: present and future}, JHEP 12
  (2016) 135.
\newblock \href {http://arxiv.org/abs/1608.01509} {\path{arXiv:1608.01509}},
  \href {https://doi.org/10.1007/JHEP12(2016)135}
  {\path{doi:10.1007/JHEP12(2016)135}}.

\bibitem{ATLAS:2017rzl}
M.~Aaboud, et~al., {Measurement of the $W$-boson mass in pp collisions at
  $\sqrt{s}=7$ TeV with the ATLAS detector}, Eur. Phys. J. C 78~(2) (2018) 110,
  [Erratum: Eur.Phys.J.C 78, 898 (2018)].
\newblock \href {http://arxiv.org/abs/1701.07240} {\path{arXiv:1701.07240}},
  \href {https://doi.org/10.1140/epjc/s10052-017-5475-4}
  {\path{doi:10.1140/epjc/s10052-017-5475-4}}.

\bibitem{LHCb:2021bjt}
R.~Aaij, et~al., {Measurement of the W boson mass}, JHEP 01 (2022) 036.
\newblock \href {http://arxiv.org/abs/2109.01113} {\path{arXiv:2109.01113}},
  \href {https://doi.org/10.1007/JHEP01(2022)036}
  {\path{doi:10.1007/JHEP01(2022)036}}.

\bibitem{deBlas:2021wap}
J.~de~Blas, M.~Ciuchini, E.~Franco, A.~Goncalves, S.~Mishima, M.~Pierini,
  L.~Reina, L.~Silvestrini, {Global analysis of electroweak data in the
  Standard Model}, Phys. Rev. D 106~(3) (2022) 033003.
\newblock \href {http://arxiv.org/abs/2112.07274} {\path{arXiv:2112.07274}},
  \href {https://doi.org/10.1103/PhysRevD.106.033003}
  {\path{doi:10.1103/PhysRevD.106.033003}}.

\bibitem{Gao:2022wxk}
J.~Gao, D.~Liu, K.~Xie, {Understanding PDF uncertainty in $W$ boson mass
  measurements*}, Chin. Phys. C 46~(12) (2022) 123110.
\newblock \href {http://arxiv.org/abs/2205.03942} {\path{arXiv:2205.03942}},
  \href {https://doi.org/10.1088/1674-1137/ac930b}
  {\path{doi:10.1088/1674-1137/ac930b}}.

\bibitem{Martin:2022qiv}
S.~P. Martin, {Three-loop QCD corrections to the electroweak boson masses},
  Phys. Rev. D 106~(1) (2022) 013007.
\newblock \href {http://arxiv.org/abs/2203.05042} {\path{arXiv:2203.05042}},
  \href {https://doi.org/10.1103/PhysRevD.106.013007}
  {\path{doi:10.1103/PhysRevD.106.013007}}.

\bibitem{Isaacson:2022rts}
J.~Isaacson, Y.~Fu, C.~P. Yuan, {ResBos2 and the CDF W Mass Measurement} (5
  2022).
\newblock \href {http://arxiv.org/abs/2205.02788} {\path{arXiv:2205.02788}}.

\bibitem{Ellis:1982wd}
R.~K. Ellis, W.~Furmanski, R.~Petronzio, {Power Corrections to the Parton Model
  in QCD}, Nucl. Phys. B 207 (1982) 1--14.
\newblock \href {https://doi.org/10.1016/0550-3213(82)90132-8}
  {\path{doi:10.1016/0550-3213(82)90132-8}}.

\bibitem{Ellis:1982cd}
R.~K. Ellis, W.~Furmanski, R.~Petronzio, {Unraveling Higher Twists}, Nucl.
  Phys. B 212 (1983) 29.
\newblock \href {https://doi.org/10.1016/0550-3213(83)90597-7}
  {\path{doi:10.1016/0550-3213(83)90597-7}}.

\bibitem{Collins:2011zzd}
J.~Collins, {Foundations of perturbative QCD}, Vol.~32, Cambridge University
  Press, 2013.

\bibitem{Konetschny:1977bn}
W.~Konetschny, W.~Kummer, {Nonconservation of Total Lepton Number with Scalar
  Bosons}, Phys. Lett. B 70 (1977) 433--435.
\newblock \href {https://doi.org/10.1016/0370-2693(77)90407-5}
  {\path{doi:10.1016/0370-2693(77)90407-5}}.

\bibitem{Magg:1980ut}
M.~Magg, C.~Wetterich, {Neutrino Mass Problem and Gauge Hierarchy}, Phys. Lett.
  B 94 (1980) 61--64.
\newblock \href {https://doi.org/10.1016/0370-2693(80)90825-4}
  {\path{doi:10.1016/0370-2693(80)90825-4}}.

\bibitem{Cheng:1980qt}
T.~P. Cheng, L.-F. Li, {Neutrino Masses, Mixings and Oscillations in SU(2) x
  U(1) Models of Electroweak Interactions}, Phys. Rev. D 22 (1980) 2860.
\newblock \href {https://doi.org/10.1103/PhysRevD.22.2860}
  {\path{doi:10.1103/PhysRevD.22.2860}}.

\bibitem{Schechter:1980gr}
J.~Schechter, J.~W.~F. Valle, {Neutrino Masses in SU(2) x U(1) Theories}, Phys.
  Rev. D 22 (1980) 2227.
\newblock \href {https://doi.org/10.1103/PhysRevD.22.2227}
  {\path{doi:10.1103/PhysRevD.22.2227}}.

\bibitem{Lazarides:1980nt}
G.~Lazarides, Q.~Shafi, C.~Wetterich, {Proton Lifetime and Fermion Masses in an
  SO(10) Model}, Nucl. Phys. B 181 (1981) 287--300.
\newblock \href {https://doi.org/10.1016/0550-3213(81)90354-0}
  {\path{doi:10.1016/0550-3213(81)90354-0}}.

\bibitem{Mohapatra:1980yp}
R.~N. Mohapatra, G.~Senjanovic, {Neutrino Masses and Mixings in Gauge Models
  with Spontaneous Parity Violation}, Phys. Rev. D 23 (1981) 165.
\newblock \href {https://doi.org/10.1103/PhysRevD.23.165}
  {\path{doi:10.1103/PhysRevD.23.165}}.

\bibitem{Ross:1975fq}
D.~A. Ross, M.~J.~G. Veltman, {Neutral Currents in Neutrino Experiments}, Nucl.
  Phys. B 95 (1975) 135--147.
\newblock \href {https://doi.org/10.1016/0550-3213(75)90485-X}
  {\path{doi:10.1016/0550-3213(75)90485-X}}.

\bibitem{Lavoura:1993nq}
L.~Lavoura, L.-F. Li, {Making the small oblique parameters large}, Phys. Rev. D
  49 (1994) 1409--1416.
\newblock \href {http://arxiv.org/abs/hep-ph/9309262}
  {\path{arXiv:hep-ph/9309262}}, \href
  {https://doi.org/10.1103/PhysRevD.49.1409}
  {\path{doi:10.1103/PhysRevD.49.1409}}.

\bibitem{Chun:2012jw}
E.~J. Chun, H.~M. Lee, P.~Sharma, {Vacuum Stability, Perturbativity, EWPD and
  Higgs-to-diphoton rate in Type II Seesaw Models}, JHEP 11 (2012) 106.
\newblock \href {http://arxiv.org/abs/1209.1303} {\path{arXiv:1209.1303}},
  \href {https://doi.org/10.1007/JHEP11(2012)106}
  {\path{doi:10.1007/JHEP11(2012)106}}.

\bibitem{Kanemura:2022ahw}
S.~Kanemura, K.~Yagyu, {Implication of the W boson mass anomaly at CDF II in
  the Higgs triplet model with a mass difference}, Phys. Lett. B 831 (2022)
  137217.
\newblock \href {http://arxiv.org/abs/2204.07511} {\path{arXiv:2204.07511}},
  \href {https://doi.org/10.1016/j.physletb.2022.137217}
  {\path{doi:10.1016/j.physletb.2022.137217}}.

\bibitem{Heeck:2022fvl}
J.~Heeck, {W-boson mass in the triplet seesaw model}, Phys. Rev. D 106~(1)
  (2022) 015004.
\newblock \href {http://arxiv.org/abs/2204.10274} {\path{arXiv:2204.10274}},
  \href {https://doi.org/10.1103/PhysRevD.106.015004}
  {\path{doi:10.1103/PhysRevD.106.015004}}.

\bibitem{Ghosh:2022zqs}
R.~Ghosh, B.~Mukhopadhyaya, U.~Sarkar, {The $\rho$ parameter and the CDF W-mass
  anomaly: observations on the role of scalar triplets} (5 2022).
\newblock \href {http://arxiv.org/abs/2205.05041} {\path{arXiv:2205.05041}}.

\bibitem{Bahl:2022gqg}
H.~Bahl, W.~H. Chiu, C.~Gao, L.-T. Wang, Y.-M. Zhong, {Tripling down on the $W$
  boson mass}, Eur. Phys. J. C 82~(10) (2022) 944.
\newblock \href {http://arxiv.org/abs/2207.04059} {\path{arXiv:2207.04059}},
  \href {https://doi.org/10.1140/epjc/s10052-022-10934-5}
  {\path{doi:10.1140/epjc/s10052-022-10934-5}}.

\bibitem{Cheng:2022hbo}
Y.~Cheng, X.-G. He, F.~Huang, J.~Sun, Z.-P. Xing, {Electroweak precision tests
  for triplet scalars} (8 2022).
\newblock \href {http://arxiv.org/abs/2208.06760} {\path{arXiv:2208.06760}}.

\bibitem{ATLAS:2017xqs}
M.~Aaboud, et~al., {Search for doubly charged Higgs boson production in
  multi-lepton final states with the ATLAS detector using
  proton\textendash{}proton collisions at $\sqrt{s}=13\,\text {TeV}$}, Eur.
  Phys. J. C 78~(3) (2018) 199.
\newblock \href {http://arxiv.org/abs/1710.09748} {\path{arXiv:1710.09748}},
  \href {https://doi.org/10.1140/epjc/s10052-018-5661-z}
  {\path{doi:10.1140/epjc/s10052-018-5661-z}}.

\bibitem{ATLAS:2018ceg}
M.~Aaboud, et~al., {Search for doubly charged scalar bosons decaying into
  same-sign $W$ boson pairs with the ATLAS detector}, Eur. Phys. J. C 79~(1)
  (2019) 58.
\newblock \href {http://arxiv.org/abs/1808.01899} {\path{arXiv:1808.01899}},
  \href {https://doi.org/10.1140/epjc/s10052-018-6500-y}
  {\path{doi:10.1140/epjc/s10052-018-6500-y}}.

\bibitem{ATLAS:2021jol}
G.~Aad, et~al., {Search for doubly and singly charged Higgs bosons decaying
  into vector bosons in multi-lepton final states with the ATLAS detector using
  proton-proton collisions at $ \sqrt{\mathrm{s}} $ = 13 TeV}, JHEP 06 (2021)
  146.
\newblock \href {http://arxiv.org/abs/2101.11961} {\path{arXiv:2101.11961}},
  \href {https://doi.org/10.1007/JHEP06(2021)146}
  {\path{doi:10.1007/JHEP06(2021)146}}.

\bibitem{ATLAS:2022yzd}
{Search for doubly charged Higgs boson production in multi-lepton final states
  using $139\,\text{fb}^{-1}$ of proton--proton collisions at $\sqrt{s}=
  13\,\text{TeV}$ with the ATLAS detector} (2022).

\bibitem{CMS:2012dun}
S.~Chatrchyan, et~al., {A Search for a Doubly-Charged Higgs Boson in $pp$
  Collisions at $\sqrt{s}=7$ TeV}, Eur. Phys. J. C 72 (2012) 2189.
\newblock \href {http://arxiv.org/abs/1207.2666} {\path{arXiv:1207.2666}},
  \href {https://doi.org/10.1140/epjc/s10052-012-2189-5}
  {\path{doi:10.1140/epjc/s10052-012-2189-5}}.

\bibitem{CMS:2016cpz}
{Search for a doubly-charged Higgs boson with $\sqrt{s}=8~\mathrm{TeV}$ $pp$
  collisions at the CMS experiment} (2016).

\bibitem{Han:2007bk}
T.~Han, B.~Mukhopadhyaya, Z.~Si, K.~Wang, {Pair production of doubly-charged
  scalars: Neutrino mass constraints and signals at the LHC}, Phys. Rev. D 76
  (2007) 075013.
\newblock \href {http://arxiv.org/abs/0706.0441} {\path{arXiv:0706.0441}},
  \href {https://doi.org/10.1103/PhysRevD.76.075013}
  {\path{doi:10.1103/PhysRevD.76.075013}}.

\bibitem{FileviezPerez:2008wbg}
P.~Fileviez~Perez, T.~Han, G.-Y. Huang, T.~Li, K.~Wang, {Testing a Neutrino
  Mass Generation Mechanism at the LHC}, Phys. Rev. D 78 (2008) 071301.
\newblock \href {http://arxiv.org/abs/0803.3450} {\path{arXiv:0803.3450}},
  \href {https://doi.org/10.1103/PhysRevD.78.071301}
  {\path{doi:10.1103/PhysRevD.78.071301}}.

\bibitem{FileviezPerez:2008jbu}
P.~Fileviez~Perez, T.~Han, G.-y. Huang, T.~Li, K.~Wang, {Neutrino Masses and
  the CERN LHC: Testing Type II Seesaw}, Phys. Rev. D 78 (2008) 015018.
\newblock \href {http://arxiv.org/abs/0805.3536} {\path{arXiv:0805.3536}},
  \href {https://doi.org/10.1103/PhysRevD.78.015018}
  {\path{doi:10.1103/PhysRevD.78.015018}}.

\bibitem{Melfo:2011nx}
A.~Melfo, M.~Nemevsek, F.~Nesti, G.~Senjanovic, Y.~Zhang, {Type II Seesaw at
  LHC: The Roadmap}, Phys. Rev. D 85 (2012) 055018.
\newblock \href {http://arxiv.org/abs/1108.4416} {\path{arXiv:1108.4416}},
  \href {https://doi.org/10.1103/PhysRevD.85.055018}
  {\path{doi:10.1103/PhysRevD.85.055018}}.

\bibitem{Ashanujjaman:2021txz}
S.~Ashanujjaman, K.~Ghosh, {Revisiting type-II see-saw: present limits and
  future prospects at LHC}, JHEP 03 (2022) 195.
\newblock \href {http://arxiv.org/abs/2108.10952} {\path{arXiv:2108.10952}},
  \href {https://doi.org/10.1007/JHEP03(2022)195}
  {\path{doi:10.1007/JHEP03(2022)195}}.

\bibitem{Mandal:2022zmy}
S.~Mandal, O.~G. Miranda, G.~Sanchez~Garcia, J.~W.~F. Valle, X.-J. Xu, {Toward
  deconstructing the simplest seesaw mechanism}, Phys. Rev. D 105~(9) (2022)
  095020.
\newblock \href {http://arxiv.org/abs/2203.06362} {\path{arXiv:2203.06362}},
  \href {https://doi.org/10.1103/PhysRevD.105.095020}
  {\path{doi:10.1103/PhysRevD.105.095020}}.

\bibitem{Stelzer:1994ta}
T.~Stelzer, W.~F. Long, {Automatic generation of tree level helicity
  amplitudes}, Comput. Phys. Commun. 81 (1994) 357--371.
\newblock \href {http://arxiv.org/abs/hep-ph/9401258}
  {\path{arXiv:hep-ph/9401258}}, \href
  {https://doi.org/10.1016/0010-4655(94)90084-1}
  {\path{doi:10.1016/0010-4655(94)90084-1}}.

\bibitem{Alwall:2014hca}
J.~Alwall, R.~Frederix, S.~Frixione, V.~Hirschi, F.~Maltoni, O.~Mattelaer,
  H.~S. Shao, T.~Stelzer, P.~Torrielli, M.~Zaro, {The automated computation of
  tree-level and next-to-leading order differential cross sections, and their
  matching to parton shower simulations}, JHEP 07 (2014) 079.
\newblock \href {http://arxiv.org/abs/1405.0301} {\path{arXiv:1405.0301}},
  \href {https://doi.org/10.1007/JHEP07(2014)079}
  {\path{doi:10.1007/JHEP07(2014)079}}.

\bibitem{Buckley:2021neu}
A.~Buckley, et~al., {Testing new physics models with global comparisons to
  collider measurements: the Contur toolkit}, SciPost Phys. Core 4 (2021) 013.
\newblock \href {http://arxiv.org/abs/2102.04377} {\path{arXiv:2102.04377}},
  \href {https://doi.org/10.21468/SciPostPhysCore.4.2.013}
  {\path{doi:10.21468/SciPostPhysCore.4.2.013}}.

\bibitem{Butterworth:2016sqg}
J.~M. Butterworth, D.~Grellscheid, M.~Kr\"amer, B.~Sarrazin, D.~Yallup,
  {Constraining new physics with collider measurements of Standard Model
  signatures}, JHEP 03 (2017) 078.
\newblock \href {http://arxiv.org/abs/1606.05296} {\path{arXiv:1606.05296}},
  \href {https://doi.org/10.1007/JHEP03(2017)078}
  {\path{doi:10.1007/JHEP03(2017)078}}.

\bibitem{Altakach:2021lkq}
M.~M. Altakach, J.~M. Butterworth, T.~Je\v{z}o, M.~Klasen, I.~Schienbein,
  {Probing a leptophobic top-colour model with cross section measurements and
  precise signal and background predictions: a case study} (11 2021).
\newblock \href {http://arxiv.org/abs/2111.15406} {\path{arXiv:2111.15406}}.

\bibitem{Fuks:2019clu}
B.~Fuks, M.~Nemev\v{s}ek, R.~Ruiz, {Doubly Charged Higgs Boson Production at
  Hadron Colliders}, Phys. Rev. D 101~(7) (2020) 075022.
\newblock \href {http://arxiv.org/abs/1912.08975} {\path{arXiv:1912.08975}},
  \href {https://doi.org/10.1103/PhysRevD.101.075022}
  {\path{doi:10.1103/PhysRevD.101.075022}}.

\bibitem{Pich:1984uoh}
A.~Pich, A.~Santamaria, J.~Bernabeu, {MU- ---\ensuremath{>} E- + GAMMA DECAY IN
  THE SCALAR TRIPLET MODEL}, Phys. Lett. B 148 (1984) 229--233.
\newblock \href {https://doi.org/10.1016/0370-2693(84)91644-7}
  {\path{doi:10.1016/0370-2693(84)91644-7}}.

\bibitem{Chakrabortty:2012vp}
J.~Chakrabortty, P.~Ghosh, W.~Rodejohann, {Lower Limits on $\mu \to e \gamma$
  from New Measurements on $U_{e3}$}, Phys. Rev. D 86 (2012) 075020.
\newblock \href {http://arxiv.org/abs/1204.1000} {\path{arXiv:1204.1000}},
  \href {https://doi.org/10.1103/PhysRevD.86.075020}
  {\path{doi:10.1103/PhysRevD.86.075020}}.

\bibitem{Cheng:2022jyi}
Y.~Cheng, X.-G. He, Z.-L. Huang, M.-W. Li, {Type-II seesaw triplet scalar
  effects on neutrino trident scattering}, Phys. Lett. B 831 (2022) 137218.
\newblock \href {http://arxiv.org/abs/2204.05031} {\path{arXiv:2204.05031}},
  \href {https://doi.org/10.1016/j.physletb.2022.137218}
  {\path{doi:10.1016/j.physletb.2022.137218}}.

\bibitem{Akeroyd:2005gt}
A.~G. Akeroyd, M.~Aoki, {Single and pair production of doubly charged Higgs
  bosons at hadron colliders}, Phys. Rev. D 72 (2005) 035011.
\newblock \href {http://arxiv.org/abs/hep-ph/0506176}
  {\path{arXiv:hep-ph/0506176}}, \href
  {https://doi.org/10.1103/PhysRevD.72.035011}
  {\path{doi:10.1103/PhysRevD.72.035011}}.

\bibitem{delAguila:2008cj}
F.~del Aguila, J.~A. Aguilar-Saavedra, {Distinguishing seesaw models at LHC
  with multi-lepton signals}, Nucl. Phys. B 813 (2009) 22--90.
\newblock \href {http://arxiv.org/abs/0808.2468} {\path{arXiv:0808.2468}},
  \href {https://doi.org/10.1016/j.nuclphysb.2008.12.029}
  {\path{doi:10.1016/j.nuclphysb.2008.12.029}}.

\bibitem{FileviezPerez:2009hdc}
P.~Fileviez~Perez, T.~Han, T.~Li, {Testability of Type I Seesaw at the CERN
  LHC: Revealing the Existence of the B-L Symmetry}, Phys. Rev. D 80 (2009)
  073015.
\newblock \href {http://arxiv.org/abs/0907.4186} {\path{arXiv:0907.4186}},
  \href {https://doi.org/10.1103/PhysRevD.80.073015}
  {\path{doi:10.1103/PhysRevD.80.073015}}.

\bibitem{Akeroyd:2011zza}
A.~G. Akeroyd, H.~Sugiyama, {Production of doubly charged scalars from the
  decay of singly charged scalars in the Higgs Triplet Model}, Phys. Rev. D 84
  (2011) 035010.
\newblock \href {http://arxiv.org/abs/1105.2209} {\path{arXiv:1105.2209}},
  \href {https://doi.org/10.1103/PhysRevD.84.035010}
  {\path{doi:10.1103/PhysRevD.84.035010}}.

\bibitem{Nemevsek:2016enw}
M.~Nemev\v{s}ek, F.~Nesti, J.~C. Vasquez, {Majorana Higgses at colliders}, JHEP
  04 (2017) 114.
\newblock \href {http://arxiv.org/abs/1612.06840} {\path{arXiv:1612.06840}},
  \href {https://doi.org/10.1007/JHEP04(2017)114}
  {\path{doi:10.1007/JHEP04(2017)114}}.

\bibitem{Cai:2017mow}
Y.~Cai, T.~Han, T.~Li, R.~Ruiz, {Lepton Number Violation: Seesaw Models and
  Their Collider Tests}, Front. in Phys. 6 (2018) 40.
\newblock \href {http://arxiv.org/abs/1711.02180} {\path{arXiv:1711.02180}},
  \href {https://doi.org/10.3389/fphy.2018.00040}
  {\path{doi:10.3389/fphy.2018.00040}}.

\bibitem{Padhan:2019jlc}
R.~Padhan, D.~Das, M.~Mitra, A.~Kumar~Nayak, {Probing doubly and singly charged
  Higgs bosons at the $pp$ collider HE-LHC}, Phys. Rev. D 101~(7) (2020)
  075050.
\newblock \href {http://arxiv.org/abs/1909.10495} {\path{arXiv:1909.10495}},
  \href {https://doi.org/10.1103/PhysRevD.101.075050}
  {\path{doi:10.1103/PhysRevD.101.075050}}.

\bibitem{Gluza:2020qrt}
J.~Gluza, M.~Kordiaczynska, T.~Srivastava, {Discriminating the HTM and MLRSM
  models in collider studies via doubly charged Higgs boson pair production and
  the subsequent leptonic decays}, Chin. Phys. C 45~(7) (2021) 073113.
\newblock \href {http://arxiv.org/abs/2006.04610} {\path{arXiv:2006.04610}},
  \href {https://doi.org/10.1088/1674-1137/abfe51}
  {\path{doi:10.1088/1674-1137/abfe51}}.

\bibitem{CMS:2022msk}
{Seesaw Model Searches Using Multilepton Final States at the HL-LHC} (2022).

\bibitem{Kanemura:2013vxa}
S.~Kanemura, K.~Yagyu, H.~Yokoya, {First constraint on the mass of
  doubly-charged Higgs bosons in the same-sign diboson decay scenario at the
  LHC}, Phys. Lett. B 726 (2013) 316--319.
\newblock \href {http://arxiv.org/abs/1305.2383} {\path{arXiv:1305.2383}},
  \href {https://doi.org/10.1016/j.physletb.2013.08.054}
  {\path{doi:10.1016/j.physletb.2013.08.054}}.

\bibitem{Kanemura:2014goa}
S.~Kanemura, M.~Kikuchi, K.~Yagyu, H.~Yokoya, {Bounds on the mass of
  doubly-charged Higgs bosons in the same-sign diboson decay scenario}, Phys.
  Rev. D 90~(11) (2014) 115018.
\newblock \href {http://arxiv.org/abs/1407.6547} {\path{arXiv:1407.6547}},
  \href {https://doi.org/10.1103/PhysRevD.90.115018}
  {\path{doi:10.1103/PhysRevD.90.115018}}.

\bibitem{Kanemura:2014ipa}
S.~Kanemura, M.~Kikuchi, H.~Yokoya, K.~Yagyu, {LHC Run-I constraint on the mass
  of doubly charged Higgs bosons in the same-sign diboson decay scenario}, PTEP
  2015 (2015) 051B02.
\newblock \href {http://arxiv.org/abs/1412.7603} {\path{arXiv:1412.7603}},
  \href {https://doi.org/10.1093/ptep/ptv071} {\path{doi:10.1093/ptep/ptv071}}.

\bibitem{BhupalDev:2018tox}
P.~S. Bhupal~Dev, Y.~Zhang, {Displaced vertex signatures of doubly charged
  scalars in the type-II seesaw and its left-right extensions}, JHEP 10 (2018)
  199.
\newblock \href {http://arxiv.org/abs/1808.00943} {\path{arXiv:1808.00943}},
  \href {https://doi.org/10.1007/JHEP10(2018)199}
  {\path{doi:10.1007/JHEP10(2018)199}}.

\bibitem{Antusch:2018svb}
S.~Antusch, O.~Fischer, A.~Hammad, C.~Scherb, {Low scale type II seesaw:
  Present constraints and prospects for displaced vertex searches}, JHEP 02
  (2019) 157.
\newblock \href {http://arxiv.org/abs/1811.03476} {\path{arXiv:1811.03476}},
  \href {https://doi.org/10.1007/JHEP02(2019)157}
  {\path{doi:10.1007/JHEP02(2019)157}}.

\bibitem{ATLAS:2021tyg}
{Combination of searches for non-resonant and resonant Higgs boson pair
  production in the $b\bar{b}\gamma\gamma$, $b\bar{b}\tau^{+}\tau^{-}$ and
  $b\bar{b}b\bar{b}$ decay channels using $pp$ collisions at $\sqrt{s}$ = 13
  TeV with the ATLAS detector} (2021).

\bibitem{Peskin:1990zt}
M.~E. Peskin, T.~Takeuchi, {A New constraint on a strongly interacting Higgs
  sector}, Phys. Rev. Lett. 65 (1990) 964--967.
\newblock \href {https://doi.org/10.1103/PhysRevLett.65.964}
  {\path{doi:10.1103/PhysRevLett.65.964}}.

\bibitem{Peskin:1991sw}
M.~E. Peskin, T.~Takeuchi, {Estimation of oblique electroweak corrections},
  Phys. Rev. D 46 (1992) 381--409.
\newblock \href {https://doi.org/10.1103/PhysRevD.46.381}
  {\path{doi:10.1103/PhysRevD.46.381}}.

\bibitem{ParticleDataGroup:2020ssz}
P.~A. Zyla, et~al., {Review of Particle Physics}, PTEP 2020~(8) (2020) 083C01.
\newblock \href {https://doi.org/10.1093/ptep/ptaa104}
  {\path{doi:10.1093/ptep/ptaa104}}.

\bibitem{deBlas:2022hdk}
J.~de~Blas, M.~Pierini, L.~Reina, L.~Silvestrini, {Impact of the recent
  measurements of the top-quark and W-boson masses on electroweak precision
  fits} (4 2022).
\newblock \href {http://arxiv.org/abs/2204.04204} {\path{arXiv:2204.04204}}.

\bibitem{Maksymyk:1993zm}
I.~Maksymyk, C.~P. Burgess, D.~London, {Beyond S, T and U}, Phys. Rev. D 50
  (1994) 529--535.
\newblock \href {http://arxiv.org/abs/hep-ph/9306267}
  {\path{arXiv:hep-ph/9306267}}, \href
  {https://doi.org/10.1103/PhysRevD.50.529}
  {\path{doi:10.1103/PhysRevD.50.529}}.

\bibitem{Binosi:2008ig}
D.~Binosi, J.~Collins, C.~Kaufhold, L.~Theussl, {JaxoDraw: A Graphical user
  interface for drawing Feynman diagrams. Version 2.0 release notes}, Comput.
  Phys. Commun. 180 (2009) 1709--1715.
\newblock \href {http://arxiv.org/abs/0811.4113} {\path{arXiv:0811.4113}},
  \href {https://doi.org/10.1016/j.cpc.2009.02.020}
  {\path{doi:10.1016/j.cpc.2009.02.020}}.

\bibitem{Lu:2022bgw}
C.-T. Lu, L.~Wu, Y.~Wu, B.~Zhu, {Electroweak precision fit and new physics in
  light of the W boson mass}, Phys. Rev. D 106~(3) (2022) 035034.
\newblock \href {http://arxiv.org/abs/2204.03796} {\path{arXiv:2204.03796}},
  \href {https://doi.org/10.1103/PhysRevD.106.035034}
  {\path{doi:10.1103/PhysRevD.106.035034}}.

\bibitem{Asadi:2022xiy}
P.~Asadi, C.~Cesarotti, K.~Fraser, S.~Homiller, A.~Parikh, {Oblique Lessons
  from the $W$ Mass Measurement at CDF II} (4 2022).
\newblock \href {http://arxiv.org/abs/2204.05283} {\path{arXiv:2204.05283}}.

\bibitem{Bierlich:2019rhm}
C.~Bierlich, et~al., {Robust Independent Validation of Experiment and Theory:
  Rivet version 3}, SciPost Phys. 8 (2020) 026.
\newblock \href {http://arxiv.org/abs/1912.05451} {\path{arXiv:1912.05451}},
  \href {https://doi.org/10.21468/SciPostPhys.8.2.026}
  {\path{doi:10.21468/SciPostPhys.8.2.026}}.

\bibitem{Maguire:2017ypu}
E.~Maguire, L.~Heinrich, G.~Watt, {HEPData: a repository for high energy
  physics data}, J. Phys. Conf. Ser. 898~(10) (2017) 102006.
\newblock \href {http://arxiv.org/abs/1704.05473} {\path{arXiv:1704.05473}},
  \href {https://doi.org/10.1088/1742-6596/898/10/102006}
  {\path{doi:10.1088/1742-6596/898/10/102006}}.

\bibitem{Christensen:2008py}
N.~D. Christensen, C.~Duhr, {FeynRules - Feynman rules made easy}, Comput.
  Phys. Commun. 180 (2009) 1614--1641.
\newblock \href {http://arxiv.org/abs/0806.4194} {\path{arXiv:0806.4194}},
  \href {https://doi.org/10.1016/j.cpc.2009.02.018}
  {\path{doi:10.1016/j.cpc.2009.02.018}}.

\bibitem{Alloul:2013bka}
A.~Alloul, N.~D. Christensen, C.~Degrande, C.~Duhr, B.~Fuks, {FeynRules 2.0 - A
  complete toolbox for tree-level phenomenology}, Comput. Phys. Commun. 185
  (2014) 2250--2300.
\newblock \href {http://arxiv.org/abs/1310.1921} {\path{arXiv:1310.1921}},
  \href {https://doi.org/10.1016/j.cpc.2014.04.012}
  {\path{doi:10.1016/j.cpc.2014.04.012}}.

\bibitem{Degrande:2011ua}
C.~Degrande, C.~Duhr, B.~Fuks, D.~Grellscheid, O.~Mattelaer, T.~Reiter, {UFO -
  The Universal FeynRules Output}, Comput. Phys. Commun. 183 (2012) 1201--1214.
\newblock \href {http://arxiv.org/abs/1108.2040} {\path{arXiv:1108.2040}},
  \href {https://doi.org/10.1016/j.cpc.2012.01.022}
  {\path{doi:10.1016/j.cpc.2012.01.022}}.

\bibitem{NNPDF:2014otw}
R.~D. Ball, et~al., {Parton distributions for the LHC Run II}, JHEP 04 (2015)
  040.
\newblock \href {http://arxiv.org/abs/1410.8849} {\path{arXiv:1410.8849}},
  \href {https://doi.org/10.1007/JHEP04(2015)040}
  {\path{doi:10.1007/JHEP04(2015)040}}.

\bibitem{Artoisenet:2012st}
P.~Artoisenet, R.~Frederix, O.~Mattelaer, R.~Rietkerk, {Automatic
  spin-entangled decays of heavy resonances in Monte Carlo simulations}, JHEP
  03 (2013) 015.
\newblock \href {http://arxiv.org/abs/1212.3460} {\path{arXiv:1212.3460}},
  \href {https://doi.org/10.1007/JHEP03(2013)015}
  {\path{doi:10.1007/JHEP03(2013)015}}.

\bibitem{Alwall:2014bza}
J.~Alwall, C.~Duhr, B.~Fuks, O.~Mattelaer, D.~G. \"Ozt\"urk, C.-H. Shen,
  {Computing decay rates for new physics theories with FeynRules and MadGraph
  5\_aMC@NLO}, Comput. Phys. Commun. 197 (2015) 312--323.
\newblock \href {http://arxiv.org/abs/1402.1178} {\path{arXiv:1402.1178}},
  \href {https://doi.org/10.1016/j.cpc.2015.08.031}
  {\path{doi:10.1016/j.cpc.2015.08.031}}.

\bibitem{Bierlich:2022pfr}
C.~Bierlich, et~al., {A comprehensive guide to the physics and usage of PYTHIA
  8.3} (3 2022).
\newblock \href {http://arxiv.org/abs/2203.11601} {\path{arXiv:2203.11601}},
  \href {https://doi.org/10.21468/SciPostPhysCodeb.8}
  {\path{doi:10.21468/SciPostPhysCodeb.8}}.

\bibitem{Skands:2014pea}
P.~Skands, S.~Carrazza, J.~Rojo, {Tuning PYTHIA 8.1: the Monash 2013 Tune},
  Eur. Phys. J. C 74~(8) (2014) 3024.
\newblock \href {http://arxiv.org/abs/1404.5630} {\path{arXiv:1404.5630}},
  \href {https://doi.org/10.1140/epjc/s10052-014-3024-y}
  {\path{doi:10.1140/epjc/s10052-014-3024-y}}.

\bibitem{ATLAS:2021kog}
G.~Aad, et~al., {Measurements of differential cross-sections in four-lepton
  events in 13 TeV proton-proton collisions with the ATLAS detector}, JHEP 07
  (2021) 005.
\newblock \href {http://arxiv.org/abs/2103.01918} {\path{arXiv:2103.01918}},
  \href {https://doi.org/10.1007/JHEP07(2021)005}
  {\path{doi:10.1007/JHEP07(2021)005}}.

\bibitem{ATLAS:2016jeu}
M.~Aaboud, et~al., {Search for triboson $W^{\pm }W^{\pm }W^{\mp }$ production
  in $pp$ collisions at $\sqrt{s}=8$ $\text {TeV}$ with the ATLAS detector},
  Eur. Phys. J. C 77~(3) (2017) 141.
\newblock \href {http://arxiv.org/abs/1610.05088} {\path{arXiv:1610.05088}},
  \href {https://doi.org/10.1140/epjc/s10052-017-4692-1}
  {\path{doi:10.1140/epjc/s10052-017-4692-1}}.

\bibitem{ATLAS:2021jgw}
G.~Aad, et~al., {Measurements of $W^+W^-+\ge 1~$jet production cross-sections
  in $pp$ collisions at $\sqrt{s}=13~$TeV with the ATLAS detector}, JHEP 06
  (2021) 003.
\newblock \href {http://arxiv.org/abs/2103.10319} {\path{arXiv:2103.10319}},
  \href {https://doi.org/10.1007/JHEP06(2021)003}
  {\path{doi:10.1007/JHEP06(2021)003}}.

\bibitem{Sherpa:2019gpd}
E.~Bothmann, et~al., {Event Generation with Sherpa 2.2}, SciPost Phys. 7~(3)
  (2019) 034.
\newblock \href {http://arxiv.org/abs/1905.09127} {\path{arXiv:1905.09127}},
  \href {https://doi.org/10.21468/SciPostPhys.7.3.034}
  {\path{doi:10.21468/SciPostPhys.7.3.034}}.

\end{thebibliography}
\end{document}